\documentclass[a4paper,11pt]{article}
\usepackage{pos}
\usepackage{bm}
\usepackage{bbm}
\usepackage{bbold}
\usepackage{braket}
\usepackage{slashed}
\usepackage{multirow}
\usepackage{caption}
\usepackage{subcaption}
\usepackage{graphicx}

\title{One Born Oppenheimer Effective Theory to rule all Exotics}

\author*[a]{Abhishek Mohapatra}

\affiliation[a]{Technical University of Munich,
TUM School of Natural Sciences, Physics Department,   
James-Franck-Str.~1, 85748 Garching, Germany.}

\emailAdd{abhishek.mohapatra@tum.de}

\abstract{The discovery of XYZ exotic states in the hadronic sector with two heavy quarks constitute one of the most important open problems in particle theory. In this work, we demonstrate that the QCD derived Born-Oppenheimer effective field theory (BOEFT) framework, provides a systematic framework to describe exotic states of arbitrary composition.  The BOEFT  construction is based on  QCD taking into consideration the scale separation and symmetry of systems containing two
heavy quarks. We show the results for the coupled channel Schrödinger equations incorporating  nonadiabatic mixing terms that describes exotic states for arbitrary quantum number of the light degrees of freedom: hybrids, tetraquarks, pentaquarks, doubly heavy baryons, and quarkonia at leading order. Additionally, we present the results of the predicted spin-symmetry multiplets, and the expression of the nonperturbative gauge-invariant correlators to be computed on lattice that are required for BOEFT: static energies, generalized Wilson loops, gluelumps, adjoint mesons or baryons, and triplet or sextext mesons or baryons.  Moreover,  for the static energies, we describe the short-distance behavior based on multipole expansion and the mixing with heavy-light thresholds at long distances based on the conserved quantum numbers.}

\FullConference{The XVIth Quark Confinement and the Hadron Spectrum Conference (QCHSC24)\\
 19-24 August, 2024\\
 Cairns Convention Centre, Cairns, Queensland, Australia\\}

\begin{document}
\maketitle

\section{Introduction}

Exotic hadrons are color-neutral bound states of quarks and gluons that exhibit unconventional structures beyond the standard meson and baryon configurations \cite{GellMann:1964nj, Zweig:1964CERN, Jaffe:1975fd}. Among them, the so-called XYZ states—containing at two heavy quarks: heavy quark-antiquark pair or heavy quark pairs—have drawn significant attention due to their masses lying around or above the heavy-light meson thresholds. These states are considered exotic because they do not fit within the expected quarkonium spectra, often exhibit anomalous decay patterns, or possess quantum numbers that defy traditional quarkonium assignments. Notable examples include the $\chi_{c1}\left(3872\right)$ and charged $Z_c$ and $Z_b$ states in the $Q\bar{Q}$ sector and the doubly heavy tetraquark state  $T_{cc}(3875)^+$ in the $QQ$ sector.  Since the discovery of $\chi_{c1}(3872)$ by the Belle collaboration~\cite{Belle:2003nnu}, numerous exotic states have been observed across different experimental facilities, including B-factories (BaBar, Belle, CLEO), \( \tau \)-charm factories (CLEO-c, BESIII), and hadron colliders (CDF, D0, LHCb, ATLAS, CMS) (see reviews~\cite{Guo:2017jvc, Ali:2017jda, Olsen:2017bmm, Brambilla:2019esw, Liu:2019zoy, Chen:2022asf}). Additionally, the LHCb experiment has discovered isospin $1/2$ baryons, such as $P_{c\bar{c}}(4380)^+$, $P_{c\bar{c}}(4312)^+$, and $P_{c\bar{c}}(4440)^+$, which are strong candidates for pentaquarks containing a charm-anticharm pair and three light quarks~\cite{LHCb:2015yax, LHCb:2019kea}.  

Numerous theoretical models based on ad-hoc choices of dominant configurations and interactions with tenous connections to QCD have been proposed to explain the nature of XYZ states. Examples include quarkonium hybrids, compact tetraquarks, diquark-diquark structures, hadronic molecules, and hadroquarkonia. However, no single model has successfully described the complete pattern. In this work,  we put forward  a QCD-derived effective field theory called Born–Oppenheimer EFT (BOEFT) as a unified framework for addressing these states, without making any assumption on their configurations. The BOEFT is derived from QCD on the basis of symmetries and scale separation and results in coupled channel Schr\"odinger equations that govern the dynamics of these states. A crucial input in these equations are the static energies which are nonperturbative quantum to be computed on lattice. While BOEFT requires lattice input for certain static energies and generalized Wilson loops, the factorization inherent in the framework ensures that only a small set of universal, flavor-independent nonperturbative correlators are needed, greatly simplifying the problem. Through a rigorous (nonperturbative) matching procedure, BOEFT provides a systematic and model-independent approach to understanding exotic hadrons, offering a unified description of XYZ states and other exotic multiquark configurations. 

In these proceedings, we briefly outline the BOEFT construction of EFTs for any states with two heavy quarks and arbitrary light degrees of freedom (LDF). We also summarize some of the new  results related to tetraquark and pentaquark obtained in Ref.~\cite{Berwein:2024ztx}.


\section{Scale separation and Born-Oppenheimer effective theory}\label{sec:effective}
Our aim is to study
hadronic states made of at least two heavy quarks (heavy quark-antiquark pair or heavy quark pairs) and LDF: gluons or light quarks. The relevant energy scales in such systems are the heavy quark mass $m_Q$, the relative momentum transfer $m_Q v$, with $v\ll 1$ being the relative velocity and 
the heavy-quark binding energy  $m_Q v^2$. Such scales are hierarchically ordered, $m_Q \gg m_Qv \gg m_Qv^2$. The LDF are characterized by the nonperturbative energy scale $\Lambda_{\mathrm {QCD}}$, which implies that the typical size of the exotic state is of order $1/\Lambda_{\mathrm {QCD}}$. The typical separation between heavy quark pair scales as $1/r\sim m_Q v$. The hierarchy of scales allow us to exploit the   nonrelativistic effective field theories~\cite{ Caswell:1985ui, Bodwin:1994jh, Manohar:1997qy, Brambilla:2004jw, Berwein:2015vca, Oncala:2017hop, Brambilla:2017uyf, Berwein:2024ztx} that turn out to be instrumental for understanding these states.

Nonrelativistic QCD (NRQCD) follows from QCD by integrating out modes associated with the bottom or the charm quark mass~\mbox{\cite{Caswell:1985ui, Bodwin:1994jh,Manohar:1997qy}}.
For quarkonium states with small radii, where $m_Q v\sim 1/r \gg \Lambda_{\mathrm {QCD}}\sim m_Qv^2$, integrating out the momentum transfer scale  $m_Qv$ leads
either to weakly-coupled potential NRQCD (pNRQCD)~\cite{Brambilla:1999xf, Pineda:1997bj}, and  for quarkonium states with larger radii, where $m_Q v \sim1/r \sim \Lambda_{\mathrm {QCD}} \gg m_Qv^2$, the same procedure leads to strongly-coupled pNRQCD ~\cite{Brambilla:2000gk,Brambilla:2004jw,Pineda:2000sz}.
The contributions from gluons (LDF) of energy and momentum of order $m_Qv$ are encoded in the  potential between the two heavy quarks which are matching coefficients of pNRQCD. In  general, the potential includes both short-range Coulomb interaction and the long-range nonperturbative piece such as the Cornell potential. Exotic states such as hybrids or tetraquarks are rather extended objects for which we assume that the LDF responsible for the  binding satisfies the strongly-coupled hierarchy  $m_Q v \sim1/r \gtrsim \Lambda_{\mathrm {QCD}} \gg m_Qv^2$. This implies that    
at least parametrically, there is no
mixing between states separated by a gap of order $\Lambda_{\mathrm {QCD}}$, which enables us to exploit the Born-Oppenheimer (BO) approximation \cite{Born-Oppenheimer, Landau:1991wop}: the nonrelativistic motion of the two heavy quarks  evolves at a time scale $1/m_Qv^2$ which is much larger than the typical time scale of LDF dynamics, $1/\Lambda_{\rm QCD}$.

The Born-Oppenheimer (BO) approximation has long been proposed as a powerful framework for describing systems containing at least two heavy quarks, particularly in the context of hybrid states \cite{Griffiths:1983ah, Juge:1999ie, Juge:2002br, Braaten:2013boa, Braaten:2014qka, Braaten:2014ita, Meyer:2015eta, Alasiri:2024nue}. This approach has been systematically developed within an effective field theory (EFT) framework in \cite{Berwein:2015vca, Brambilla:2017uyf, Oncala:2017hop, Berwein:2024ztx}.
For hybrid states, semi-inclusive decays to quarkonia have been computed using BOEFT \cite{Oncala:2017hop, Brambilla:2022hhi, TarrusCastella:2021pld}, and spin effects have been incorporated through spin-dependent potentials \cite{Oncala:2017hop, Brambilla:2018pyn, Brambilla:2019jfi, Soto:2023lbh}. Notably, these spin effects appear already at order $1/m_Q$ in contrast to heavy quarkonium systems where spin-dependent effects typically arise only at order $1/m_Q^2$. Beyond hybrids, 
general applications of the BOEFT have been put forward in \cite{Brambilla:2017uyf,Soto:2020xpm, Berwein:2024ztx}. The BOEFT and BO approximation in general has been applied to  the doubly heavy baryons in   \cite{Soto:2020pfa,Soto:2021cgk,Brambilla:2005yk}, the doubly heavy $\left(QQ/\bar{Q}\bar{Q}\right)$ and hidden heavy $\left(Q\bar{Q}\right)$ tetraquarks  in \cite{Bicudo:2015kna, Bicudo:2016ooe, Bicudo:2017szl, Bicudo:2019ymo, Braaten:2020nwp, Brambilla:2024thx, Braaten:2024tbm,Bruschini:2024fyj, Hoffmann:2024hbz}  
and describing the effect of the interaction with the heavy-light threshold  in 
\cite{Bicudo:2019ymo, TarrusCastella:2022rxb,Bruschini:2023zkb,Bruschini:2023tmm, TarrusCastella:2024zps, Braaten:2024stn}. The BOEFT is thus an  extension of the strongly coupled pNRQCD to states with non-trivial LDF.
 
In BOEFT, the hadronic states with two heavy quarks are constructed as eigenstates of the Hamiltonian at a given level of the expansions. At leading order in NRQCD, the dominant contributions are of order ${\cal O}\left(1/m_Q\right)^0$ known as \textit{static approximation}. The corresponding eigenstates and eigenvalues of the static Hamiltonian are referred to as \textit{static states} and \textit{static energies} respectively.   In the static limit, the heavy quarks are static and supply a
color source, while the different LDF configurations  identify different static energies. Thus, static energies are characterized  by the  separation $r$ between the two heavy quarks, the flavor composition of the light degrees of freedom (isospin quantum number $I$ and projection $m_I$, baryon number $b$ etc), and the projections of the LDF total angular momentum $\bm K$ along the  heavy quark pair axis $\hat{\bm r}$, whose eigenvalues are given by $\lambda$. The static energies are labelled by  the  cylindrical symmetry group $D_{\infty h}$ representations $\Lambda^{\sigma}_\eta$, which we refer to as the \textit{BO quantum numbers}: $\Lambda \equiv |\lambda| = |\bm{K}\cdot\hat{\bm{r}}|$  and  for integer values is denoted by capital Greek letters: $\Sigma$, $\Pi$, $\Delta,...$ for $\Lambda=0,1,2,...$. The index $\eta$ is the $CP$ eigenvalue in case of $Q\bar{Q}$, and just parity, $P$, in case of $QQ$ and is denoted by $g = + 1$ and $u = - 1$. 
Finally, the index $\sigma=\pm 1$ is the eigenvalue of the reflection operator  with respect to a plane passing through the $\hat{\bm{r}}$ axis. 
The index $\sigma$ is explicitly written only for $\Sigma$ states, which are not degenerate with respect to the reflection symmetry.
The ground state with given BO quantum numbers is labeled $\Lambda_\eta^\sigma$;
excited states with the same quantum numbers are labeled 
$\Lambda_\eta^{\sigma\prime}$, $\Lambda_\eta^{\sigma\prime\prime}$, \ldots. 

Assuming that the LDF angular momentum operator $\bm{K}^2$ has 
eigenvalues $k(k+1)$, which restricts the projection to $\Lambda\leq k$, we introduce a compact notation for representing LDF states  
\begin{equation}
	\kappa\equiv\{k^{P[C]}, f\}\,,
	\label{label-n}
\end{equation}
where  $P[C]$ is defined to be $PC$ in case of $Q\bar{Q}$ and just $P$ in case of $QQ$ and $f$ denotes the flavor index of the LDF \footnote{For notational ease, we will explicitly mention whenever we suppress the flavor index $f$}. Note that for quarkonium state, $k^{PC}=0^{++}$. In the static limit, the heavy-quark-heavy antiquark sector and two-heavy quark sector of the Fock space is spanned 
by
\begin{align}
\vert \underline{\kappa, \lambda}; \bm{x}_1 ,\bm{x}_2 \rangle^{(0)} & = \psi^{\dagger}(\bm{x}_1) \chi (\bm{x}_2)
|\kappa, \lambda;\bm{x}_1 ,\bm{x}_2\rangle^{(0)},\qquad \forall \bm{x}_1,\bm{x}_2\,,\label{basis0}\\
\vert \underline{\kappa, \lambda}; \bm{x}_1 ,\bm{x}_2 \rangle^{(0)} & = \psi^{\dagger}(\bm{x}_1) \psi^{\dagger}(\bm{x}_2)
|\kappa, \lambda;\bm{x}_1 ,\bm{x}_2\rangle^{(0)},\qquad \forall \bm{x}_1,\bm{x}_2\,,\label{basis1}
\end{align}
where $|\underline{\kappa, \lambda}; \bm{x}_1 ,\bm{x}_2\rangle^{(0)} $ is a gauge-invariant eigenstate of the static NRQCD Hamiltonian (defined upto a phase) with static eigenenergy $E_{\Lambda^{\sigma}_{\eta}}^{(0)}(\bm{x}_1 ,\bm{x}_2)\equiv E_{\kappa, |\lambda|}^{(0)}(\bm{x}_1 ,\bm{x}_2)$ that transforms like $3_{\bm{x}_1}\otimes 3_{\bm{x}_2}^{\ast}$ or $3_{\bm{x}_1}\otimes 3_{\bm{x}_2}$  under color $SU(3)$. Note, that the  state $|\kappa, \lambda;\bm{x}_1 ,\bm{x}_2\rangle^{(0)}$  encodes the purely LDF content of the state, and it is annihilated by the heavy quark fields for any $\bm{x}$. Moreover, the static Hamiltonian $ H^{(0)}$ does not contain any heavy fermion field, which implies the state $|\kappa, \lambda;\bm{x}_1,\bm{x}_2\rangle^{(0)}$ is also an eigenstate of static NRQCD Hamiltonian with energy $E_{\kappa, |\lambda|}^{(0)}(\bm{x}_1 ,\bm{x}_2)$. The normalization of the states are defined as
\begin{equation}
^{(0)}\langle \underline{\kappa, \lambda}; \bm{x}_1 ,\bm{x}_2|\underline{\kappa^\prime, \lambda^\prime}; \bm{y}_1 ,\bm{y}_2\rangle^{(0)} =\delta_{\kappa\kappa^\prime}\delta_{\lambda\lambda^\prime}
\delta^{(3)} (\bm{x}_1 -\bm{y}_1)\delta^{(3)} (\bm{x}_2 -\bm{y}_2)\,.
\label{norm}
\end{equation}
We have made explicit the positions $\bm{x}_1$ and $\bm{x}_2$ of the heavy quark pair $\left(Q\bar{Q}, QQ\right)$, which are good quantum numbers in the static limit. The eigenenergies $E^{\left(0\right)}_{\kappa, |\lambda|}\left({\bm x}_1, {\bm x}_2\right)$ are the LDF static energies and by translational invariance $E^{\left(0\right)}_{\kappa, |\lambda|}\left({\bm x}_1, {\bm x}_2\right)= E^{\left(0\right)}_{\kappa, |\lambda|}\left(r\right)$, where ${\bm r}={\bm x}_1-{\bm x}_2$. The static energies $E^{\left(0\right)}_{\kappa, |\lambda|}(r)$  are nonperturbative quantities computed  on the lattice. In the short distance $r\rightarrow 0$ limit,  static energies $E_{\kappa, |\lambda|}^{(0)}(r)$ with the same $\kappa$ but different $\Lambda=|\lambda|$ become degenerate, leading to an enhancement of the cylindrical group $D_{\infty h}$ to the spherical group $O\left(3\right)\otimes C$ \cite{Brambilla:1999xf}.

The static states are not bound states, as the static Hamiltonian does not contain kinetic terms. When going beyond the static limit, the kinetic energy associated with the relative motion of the heavy quarks dominates over that of the center-of-mass motion, as dictated by the power counting in pNRQCD. In a first approximation, it therefore suffices to include only the relative kinetic term when determining the bound state energies. The eigenstates of the Hamiltonian in this beyond-static approximation are linear superpositions of static states with the same quantum numbers. 

To derive the BOEFT from NRQCD, which is obtained by integrating out modes of order $\Lambda_{\mathrm{QCD}}$, we utilize the nonperturbative quantum matching framework developed in \cite{Berwein:2024ztx}. This matching procedure is carried out systematically in the $1/m_Q$ expansion. Denoting the eigenstates of the full NRQCD Hamiltonian including the $1/m_Q$ kinetic term by $| \underline{\kappa, \lambda}; \bm{x}_1, \bm{x}_2 \rangle$, the explicit matching between NRQCD and BOEFT for the states is given by
\begin{equation}
	| \underline{\kappa, \lambda}; \bm{x}_1, \bm{x}_2 \rangle \to {\Psi}^{\dagger}_{\kappa\lambda}(\bm{x}_1, \bm{x}_2)|{\rm vac} \rangle,
\end{equation}
where 
$|{\rm vac} \rangle$ is the BOEFT vacuum and $\Psi_{\kappa\lambda}(\bm{x}_1 ,\bm{x}_2)$ is a color singlet composite field that annihilates 
all states associated with LDF quanutm number $\kappa$ and static energy $E^{(0)}_{\kappa,|\lambda|}$.
Keeping at order $1/m_Q$ only the relative kinetic energy, 
the BOEFT Lagrangian describing the heavy exotic states (hybrids, tetraquarks, pentaquarks  and doubly heavy baryons)  can be written as
\begin{align}
	L_{\mathrm {BOEFT}}=&\int d^3{\bm R}\int d^3{\bm r} \, \sum_{\kappa\lambda\lambda^{\prime}}\mathrm{Tr}\Bigg\{{\Psi}^{\dagger}_{\kappa\lambda}(\bm{r},\,\bm{R},\,t) \bigg[i\partial_t\,\delta_{\lambda\lambda'}-
	V_{\kappa, \lambda\lambda^\prime }(r)   \nonumber\\
	&\hspace{4 cm}+ \sum_\alpha P^{\alpha\dag}_{\kappa\lambda}\left(\theta, \varphi\right)\frac{\bm{\nabla}^2_r}{m_Q}P^{\alpha}_{\kappa\lambda^{\prime}}\left(\theta, \varphi\right)\bigg]{\Psi}_{\kappa\lambda^{\prime}}(\bm{r},\,\bm{R},\,t)\Bigg\}, 
	\label{eq:LQQ}
\end{align}
where the trace is over color, spin and isospin indices, relative coordinate $\bm{r}\equiv {\bm x}_1-{\bm x}_2$, and the center of mass coordinate $\bm{R}\equiv \left({\bm x}_1+{\bm x}_2\right)/2$. 
Here the potential $V_{\kappa\lambda\lambda^{\prime}}$ in Eq.~\eqref{eq:LQQ} is 
\begin{align}
	&V_{\kappa, \lambda\lambda^{\prime}}(r) 
	=V^{(0)}_{\kappa, \vert\lambda \vert}(r)\delta_{\lambda\lambda^{\prime}}+{\cal O}\left(1/m_Q\right)\,,
	\label{eq:VQQ}
\end{align}
where $V^{(0)}_{\kappa, \vert\lambda \vert}(r)$
is equal to  the static energy $E^{(0)}_{\kappa, \vert\lambda \vert}(r)$. The projection vectors $P^{\alpha}_{\kappa\lambda}$ project the eigenstates of $\bm{K}\cdot\bm{\hat{r}}$ with eigenvalue $\lambda$  onto the eigenstates of $K_z$ with eigenvalue $\alpha$ and the general expression in terms of Wigner D-matrices are given in \cite{Berwein:2024ztx}.

\section{Coupled Schr\"odinger equation and multiplets} \label{sec:Sch}

The static eigenstate  $\left|\underline{\kappa, \lambda};\,\bm{x}_1,\,\bm{x}_2\right\rangle^{(0)}$ forms a complete basis. For a fixed LDF configuration with quanutm number $\kappa$, any general state $\ket{X^{(N)}}$ can be written as an expansion in them (using center of mass frame ${\bm R}=0$):
\begin{align}
 \ket{X^{(N)}}&=\sum_{\lambda} \int d^3{\bm r}\, \left|\underline{\kappa, \lambda};\,\bm{r}\right\rangle^{(0)}\,\varphi^{(N)}_{\kappa\lambda}\left({\bm r}\right)=\sum_{\lambda} \int d^3{\bm r}\, \left|\bm{r}\rangle\,\otimes\,|\kappa, \lambda\right\rangle\,\varphi^{(N)}_{\kappa\lambda}\left({\bm r}\right)\,,
 \label{eq:X1}
\end{align}
where $\varphi^{(N)}_{\kappa\lambda}\left({\bm r}, {\bm R}\right)$ are the wavefunctions that can be expressed as 
\begin{equation}
\phi^{(N)}_{\kappa\lambda}(\bm{r})=\,^{(0)}\langle \underline{\kappa, \lambda};\,\bm{r} \ket{X^{(N)}}\,.
\label{eq:Psiwf}
\end{equation}
The integration over the coordinates ${\bm r}$ implies that we are going beyond the static limit and for non-static systems, the heavy quark pair positions are not good quantum numbers. In the second equality, we have separated the Hilbert space corresponding to the static heavy quark pair and LDF:  $\left|\underline{\kappa, \lambda};\,\bm{r}\right\rangle^{(0)}\equiv \left|\bm{r}\rangle\,\otimes\,|\kappa, \lambda\right\rangle$. The quantum number $N$ of the state $\ket{X^{(N)}}$ denotes all quantum numbers of the system described by the full Hamiltonian and it contains a fixed LDF quanutm number $\kappa$: $ N \equiv \{m, j, m_j, l, s\}$: $m$ is the principle quantum number,
$l(l+1)$ is the eigenvalue of $\bm{L}^2$, $\bm{L}=\bm{L}_{Q}+\bm{K}$ being the combined angular momentum, which is
sum of the orbital angular momentum ${\bm L}_Q$ of the $Q\bar{Q}$ or $QQ$ pairs and the angular momentum ${\bm K}$ of the LDF,
$s(s+1)$ is the eigenvalue of $\bm{S}^2$, ${\bm S}={\bm S_1}+{\bm S_2}$ being the spin of the $Q\bar{Q}$ or $QQ$ pairs, 
and $j(j+1)$ and $m_j$ are the eigenvalues of $\bm{J}^2$ and $J_3$ respectively, $\bm{J}=\bm{L}+\bm{S}$ being the total angular momentum. In  $r\to0$ limit, the static states are approximate eigenstates of $\bm{K}^2$. The static energies with different $\Lambda=|\lambda|$ but same $\kappa$ are degenerate at  
order $r^0$ in the multipole expansion and the difference
only appears at higher orders in $r$, where also the symmetry under $\bm{K}^2$ is broken.


The equation of motion of the BOEFT Lagrangian in Eq.~\eqref{eq:LQQ} results in the following Schr\"odinger equation for the wave-function $\phi^{(N)}_{\kappa\lambda}\left(\bm {r}\right)$ 
\begin{align}
	\sum_{\lambda}  \left[-P^{\alpha\dag}_{\kappa\lambda'}\left(\theta, \varphi\right)\frac{\bm{\nabla}^2_r}{m_Q}\,P^{\alpha}_{\kappa\lambda}\left(\theta, \varphi\right)\,+ V_{\kappa, \lambda'\lambda}\right]\phi_{\kappa\lambda}^{(N)}\left({\bm r}\right)= {\cal E}_N\, \phi_{\kappa\lambda^\prime}^{(N)}\left({\bm r}\right),
	\label{eq:Sch1}
\end{align}
where ${\cal E}_N$ is the energy of the bound state. The kinetic term in the Schr\"odinger equation  can be split into a radial and an angular (orbital) term. In  short-distance limit $r\ll\Lambda_\mathrm{QCD}^{-1}$,  the angular term which is proportional to $1/r^2$, is the most important term. Eq.~\eqref{eq:Sch1} couple the LDF to the heavy quark dynamics through the angular part of the heavy quark kinetic operator. Since the derivatives act on both the wave function and the projectors, this results in off-diagonal matrix elements of the heavy quark kinetic energy operator, known as non-adiabatic coupling terms (NACTs). The potential $V_{\kappa\lambda\lambda'}$  in Eq.~\eqref{eq:Sch1} is a potential matrix (in the $\lambda-\lambda'$ index) with static energies in the diagonal entries (see Eq.~\eqref{eq:VQQ}). Separating the radial and the angular parts, the radial Schr\"odinger equation whose orbital term  takes on a matrix form mixing these static states can be written as
\begin{align}
	\sum_{\lambda}  \left[-\frac{1}{m_Qr^2}\partial_r\,r^2\partial_r+\frac{1}{m_Qr^2}\,M_{\lambda'\lambda}+ V_{\kappa, \lambda'\lambda}\right]\psi_{\kappa\lambda}^{(N)}\left(r\right)= {\cal E}_N\, \psi_{\kappa\lambda'}^{(N)}\left(r\right),  \label{eq:Sch2}  
\end{align}
where  $\psi_{\kappa\lambda'}^{(N)}\left(r\right)$ are the radial wavefunctions and $M_{\lambda'\lambda}$ are the mixing matrices, which results from the matrix elements of $\bm{L}_Q^2$ between the  eigenstates labeled by combined angular momentum $l$:
\begin{align}
	M_{\lambda'\lambda}&=\bigl(l(l+1)-2\lambda^2+k(k+1)\bigr)\delta_{\lambda'\lambda}-\sqrt{k(k+1)-\lambda(\lambda+1)}\sqrt{l(l+1)-\lambda(\lambda+1)}\delta_{\lambda'\lambda+1}\notag\\
	&\quad-\sqrt{k(k+1)-\lambda(\lambda-1)}\sqrt{l(l+1)-\lambda(\lambda-1)}\delta_{\lambda'\lambda-1}\,,
	\label{eq:Mlambdalambdap}
\end{align}
which is valid under the assumption that $k$ is a good quantum number which is smoothly broken at large distances.
The mixing matrix $M_{\lambda'\lambda}$ have dimensions $\min\left(2k+1, 2l+1\right)$ and can be put into block-diagonal form, where each of the  blocks corresponds to either positive or negative parity states, whose details are in Ref.~\cite{Berwein:2024ztx}. Depending on the LDF configurations corresponding to quantum number $\kappa$ in Eq.~\eqref{label-n}, the Eqs.~\eqref{eq:Sch2} and \eqref{eq:Mlambdalambdap} together gives the coupled channel Schroedinger eqautions for quarkonium, hybrids, doubly heavy baryons, tetraquarks and pentaquarks. The explicit Schr\"odinger equations for different states can be found in appendix~I of Ref.~\cite{Berwein:2024ztx}. \textit{Note that the results
on the coupled equations are new for tetraquarks and pentaquarks, while they
reproduce the previous results for hybrids and doubly heavy baryons} \cite{Berwein:2015vca, Oncala:2017hop, Soto:2020pfa}.

Quarkonium and quarkonium hybrids are states without light quarks and have gluons as LDF. We refer the reader to Ref.~\cite{Brambilla:2004jw} for discussions on quarkonium mulitplet and  Refs.~\cite{Berwein:2015vca, Oncala:2017hop} for discussions on quarkonium hybrids mulitplets.  On the contrary states like tetraquark, pentaquark and doubly heavy baryons have light quarks as LDF, which brings in additional quanutm numbers assciated with flavor (see Eq.~\eqref{label-n}). Restricting to light quarks $u$ and $d$, the flavor quanutm numbers are the isospin $I$ and the projection $m_I$. The tetraquarks have integer isospin quantum number while doubly heavy baryon and pentaquarks have half-integer isospin quantum number.

For the $Q\bar{Q}q\bar{q}$ tetraquark, both heavy and light quark  and antiquark  are distinguishable, allowing various spin, isospin, and color combinations. The combination of the color quantum numbers of the quark and antiquark results in both color singlet and
color octet configurations. The color-singlet combination corresponds to a quarkonium  accompanied by a light hadron, such as pions. However, lattice calculations \cite{Alberti:2016dru, Prelovsek:2019ywc} have shown that these states are not sufficiently bound to form a four-quark system. Instead, we consider the color-octet $\left(q\bar{q}\right)_8$—referred to as the adjoint meson—which combines with $\left(Q\bar{Q}\right)_8$ to form a color-singlet $Q\bar{Q}q\bar{q}$ tetraquark. Without orbital excitation, the lowest $q\bar{q}$ states are $k^{PC}=0^{-+}$ and $1^{--}$ which correspond to static energies  $\Sigma_u^-$ and $\{\Sigma_g^{+\prime},\Pi_g\}$, respectively. In Table~\ref{tab:QQbarqqbar}, we show the results of $J^{PC}$ multiplets of the lowest  $Q\bar{Q}q\bar{q}$ tetraquark states. The lowest tetraquark spin-symmetry multiplet in the $\{\Sigma_u^-\}$  has  $J^{PC}=\{0^{++}, 1^{+-}\}$ corresponding to the $l=0$. 
The lowest tetraquark spin-symmetry multiplet in the $\{\Sigma_g^{+\prime}, \Pi_g\}$  has  $J^{PC}=\{1^{+-},(0,1,2)^{++}\}$ corresponding to the first $l=1$ mixed state coming from the coupled Schr\"odinger equation (see Eqs.~\eqref{eq:Sch2} and \eqref{eq:Mlambdalambdap}).  We refer to Ref.~\cite{Berwein:2024ztx} regarding details on derivation of parity and charge-conjugation quantum numbers.

\begin{table}[h!]
\centering
\resizebox{0.70\columnwidth}{!}{%
	\begin{tabular}{||c|c|c||c|c|c||}
		\hline\hline
		\multirow{2}{*}{\hspace{2pt}$\begin{array}{c} Q\bar{Q}\\\text{color state}\end{array}$\hspace{2pt}} & \multirow{2}{*}{\hspace{2pt}$\begin{array}{c}\text{$q\bar{q}$ spin}\\k^{PC}\end{array}$\hspace{2pt}} & \multirow{2}{*}{\hspace{2pt} $\begin{array}{c} \text{BO quantum \#}\\\Lambda^\sigma_\eta\end{array}$\hspace{2pt}}&\multirow{2}{*}{\hspace{2pt} $l$\hspace{2pt}}& \multirow{2}{*}{\hspace{2pt} $\begin{array}{c}J^{PC}\\\{S=0, S=1\}\end{array}$\hspace{2pt}}& \multirow{2}{*}{\hspace{2pt}Multiplets\hspace{2pt}}\\
		& & & & &  \\
		\hline\hline
		\multirow{6}{*} {\hspace{2pt}$\begin{array}{c} \text{Octet}\\\mathbf{8}\end{array}$\hspace{2pt}}
		&\multirow{3}{*}{$0^{-+}$} & \multirow{3}{*}{$\Sigma_u^-$}  & \hspace{2pt}$0$\hspace{2pt} & \hspace{2pt}$\{0^{++}, 1^{+-}\}$\hspace{2pt}
		&\hspace{2pt}$T_1^0$\hspace{2pt}\\
		\cline{4-6}
		& & & \hspace{2pt}$1$\hspace{2pt} & \hspace{2pt}$\{1^{--}, \left(0, 1, 2\right)^{-+}\}$\hspace{2pt}&\hspace{2pt}$T_2^0$\hspace{2pt}\\
		\cline{2-5}\cline{2-6}
		&\multirow{4}{*}{$1^{--}$} & ${\Sigma_g^{+\prime},\Pi_g}$  & \hspace{2pt}$1$\hspace{2pt} & \hspace{2pt}$\{1^{+-}, (0,1,2)^{++}\}$\hspace{2pt}&\hspace{2pt}$T_1^1$\hspace{2pt} \\
		\cline{3-6}
		& & ${\Sigma_g^{+\prime}}$  & \hspace{2pt}$0$\hspace{2pt} & \hspace{2pt}$\{0^{-+},  1^{--}\}$\hspace{2pt}&\hspace{2pt}$T_2^1$\hspace{2pt} \\
		\cline{3-6}
		& & ${\Pi_g}$  & \hspace{2pt}$1$\hspace{2pt} & \hspace{2pt}$\{1^{-+},  (0,1,2)^{--}\}$\hspace{2pt}&\hspace{2pt}$T_3^1$\hspace{2pt} \\
		\cline{3-6}
		\hline\hline
	\end{tabular}}
	\caption{$J^{PC}$ multiplets for the lowest $Q\bar{Q}q\bar{q}$ tetraquarks \cite{Berwein:2024ztx}, with corresponding BO quantum numbers listed in the third column.  Multiplets are labeled as $T_i^k$ in the last column in order of increasing energy. The different spin combinations $\{S=0, S=1\}$
        form degenerate multiplets in absence of spin interactions. We denote the tetraquark BO quantum number as $\Sigma_g^{+\prime}$ given quarkonium BO quantum number is $\Sigma_g^+$.}
	\label{tab:QQbarqqbar}
\end{table}

For the doubly heavy $QQ\bar{q}\bar{q}$ tetraquark, the indistinguishability of the two heavy quarks and two light antiquarks constrains their quantum numbers via the Pauli exclusion principle. The quark (antiquark) color combinations form both antitriplet (triplet) and sextet (antisextet) configurations. We consider the color-triplet $\left(\bar{q}\bar{q}\right)_3$—referred to as triplet meson—which pairs with $\left(QQ\right)_{\bar{3}}$,  and the color-antisextet $\left(\bar{q}\bar{q}\right)_{\bar{6}}$—referred to as sextet meson—which pairs with $\left(QQ\right)_6$ to form a color-singlet hadrons. However, since the $\left(QQ\right)_6$ potential is repulsive at short distances, these states are unlikely to be low-lying. The two light antiquarks (quarks) can form spin and isospin singlets or triplets, but the Pauli exclusion principle restricts allowed combinations of spin, isospin, and color. The color-triplet and (iso)spin singlet are antisymmetric, while the color-antisextet and (iso)spin triplet are symmetric. This permits $(I=0, k=0)$ and $(I=1, k=1)$ states in the triplet sector, and $(I=0, k=1)$ and $(I=1, k=0)$ in the antisextet sector. Table~\ref{tab:QQqq} presents the $J^{PC}$ multiplets of the lowest $QQ\bar{q}\bar{q}$ states. Assuming that the mixed states in $\{\Sigma_g^{-}, \Pi_g\}$ have lower energies due to coupled equations, only one candidate remains: $J^{P}=(0,1,2)^+$ multiplet with $(I=1, k=1, S_Q=1, l=1)$ in the color-antitriplet sector. We refer to Ref.~\cite{Berwein:2024ztx} regarding derivation of parity quantum number.

\begin{table}[h!]
\centering
\resizebox{0.70\columnwidth}{!}{%
	\begin{tabular}{||c|c|c|c|c||c|c||}
		\hline\hline
		\multirow{2}{*}{\hspace{2pt}$\begin{array}{c} QQ\\\text{ color state}\end{array}$\hspace{2pt}} & \multirow{2}{*}{\hspace{2pt}$\begin{array}{c} \text{$\bar{q}\bar{q}$ spin}\\ k^{P}\end{array}$\hspace{2pt}} & \multirow{2}{*}{\hspace{2pt} $\begin{array}{c} \text{BO quantum \#}\\\Lambda^\sigma_\eta\end{array}$\hspace{2pt}} & \multirow{2}{*}{\hspace{2pt}$\begin{array}{c} \text{Isospin}\\I\end{array}$ \hspace{2pt}}& \multirow{2}{*}{\hspace{2pt} $l$\hspace{2pt}} & \multicolumn{2}{|c||}{$J^{P}$}\\
		\cline{6-7}
		& & & & & \hspace{2pt}$S=0$\hspace{2pt} & \hspace{2pt}$S=1$\hspace{2pt} \\
		\hline\hline
		\multirow{4}{*}{\hspace{2pt}$\begin{array}{c} \text{Antitriplet}\\\bar{\mathbf{3}}\end{array}$\hspace{2pt}} &\multirow{2}{*}{$0^+$} & \multirow{2}{*}{${\Sigma_g^+}$} & \multirow{2}{*}{$0$}& $0$ & \hspace{2pt} --- & \hspace{2pt}$1^+$\hspace{2pt} \\
		\cline{5-7}
		& & & & 1 &\hspace{2pt}$1^-$\hspace{2pt}&---\\
		\cline{2-7}
		& \multirow{2}{*}{$1^+$} & \multirow{2}{*}{${\Sigma_g^-, \Pi_g}$} & \multirow{2}{*}{$1$}& $0$ &  \hspace{2pt}$0^-$\hspace{2pt}& --- \\
		\cline{5-7}
		& & & & 1 & $1^{-}$&\hspace{2pt}$\left(0, 1, 2\right)^+$\hspace{2pt}\\
		\cline{5-7}
		\hline\hline
	\end{tabular}}
	\caption{$J^{PC}$ multiplets for the lowest $QQ\bar{q}\bar{q}$ tetraquarks \cite{Berwein:2024ztx} considering only the color antitriplet configuration $\left(QQ\right)_{\bar{3}}$, with corresponding BO quantum numbers listed in the third column. Since, the $\left(QQ\right)_{\bar{3}}$ color antitriplet potential is attractive while $\left(QQ\right)_{6}$ color sextet potential is repulsive, we expect the states with $\left(QQ\right)_{\bar{3}}$ configuration to be lower in energy. The dashed entry in last column refers to state not allowed due to the Pauli exclusion principle. }
	\label{tab:QQqq}
\end{table}

Half-integer isospin is encountered exclusively in configurations with a non-zero baryon number. The simplest of such configurations are $QQq$ baryons, $Q\bar{Q}qqq$ and $QQqq\bar{q}$ pentaquarks. Since,  we are not aware of a generally used notation for half-integer representations of $D_{\infty h}$, we choose to label the BO quantum number as $\left(k\right)_\eta$, where $\eta$ denotes parity. For ground state $QQq$ multiplets, we refer to Table.~II in Ref.~\cite{Soto:2020pfa}. For $Q\bar{Q}qqq$ pentaquark, the three light quarks have either color singlet or color octet configurations to combine with the corresponding color states of the $Q\bar{Q}$ pair to form a color-neutral state. We again only consider the $\left(qqq\right)_8$ configuration-referred to as adjoint baryon. The light quarks can form spin and isospin doublet or quartet combinations (i.e., $I=1/2$ or $I=3/2$), however, due to Pauli exclusion principle, the whole color-spin-isospin combination (ignoring the orbital excitations) needs to be fully antisymmetric regarding particle exchange. So, in the color-octet sector any combination of spin and isospin is allowed except for both being quartets $\left(I=3/2, k=3/2\right)$. In Table.~\ref{tab:QQbarpenta}, we show the results of $J^{P}$ multiplets of the lowest $Q\bar{Q}qqq$ pentaquark states. For $QQqq\bar{q}$ pentaquark, the Pauli principle constrains the quanutm numbers of the two heavy quarks and two light quarks. Again, the color antitriplet from the heavy quarks can couple to the triplet of the light quarks or the heavy sextet to the light antisextet, however, the triplet configuration is assumed to be lower in energy. Note that Pauli exclusion principle in $QQqq\bar{q}$ is less restrictive. So,  in the color-triplet sector, possible spin-isospin combinations include four doublet-doublet, two doublet-quartet, two quartet-doublet, and one quartet-quartet configuration. In the color-antisextet sector, there are two doublet-doublet states, along with one each of doublet-quartet, quartet-doublet, and quartet-quartet configurations. We refer to Ref.~\cite{Berwein:2024ztx} for the $J^P$ spin-symmetry mulitplets of $QQqq\bar{q}$ pentaquark states.
\begin{table}[h!]
	\begin{tabular}{||c|c|c||c|c||}
		\hline\hline
		\multirow{2}{*}{\hspace{2pt}$\begin{array}{c} Q\bar{Q}\\\text{color state}\end{array}$\hspace{2pt}} & \multirow{2}{*}{\hspace{2pt}$\begin{array}{c}\text{Light spin}\\k^{P}\end{array}$\hspace{2pt}} & \multirow{2}{*}{\hspace{2pt} $\begin{array}{c}\text{BO quantum \#}\\D_{\infty h}\end{array}$\hspace{2pt}} &\multirow{2}{*}{\hspace{2pt} $l$\hspace{2pt}}& \multirow{2}{*}{\hspace{2pt} $\begin{array}{c}J^{P}\\\{S_Q=0, S_Q=1\}\end{array}$\hspace{2pt}}\\
		& &  &  &  \\
		\hline\hline
		\multirow{2}{*} {\hspace{2pt}$\begin{array}{c} \text{Octet}\\\mathbf{8}\end{array}$\hspace{2pt}} &\multirow{1}{*}{$(1/2)^{+}$}& \multirow{1}{*}{$\left(1/2\right)_g$}\hspace{2pt}  & \hspace{2pt}$1/2$\hspace{2pt} & \hspace{2pt}$\{1/2^{-}, \left(1/2, 3/2\right)^-\}$\hspace{2pt}
		\\
		\cline{2-5}
		&\multirow{1}{*}{$(3/2)^{+}$}  & \multirow{1}{*}\{$\left(1/2\right)_g^{\prime}$, $\left(3/2\right)_g$\}\hspace{2pt} & \hspace{2pt}$3/2$\hspace{2pt} & \hspace{2pt}$\{3/2^{-}, (1/2, 3/2, 5/2)^{-}\}$\hspace{2pt}\\
		\hline\hline
	\end{tabular}
	\caption{$J^{P}$ multiplets for the lowest $Q\bar{Q}qqq$ pentaquarks \cite{Berwein:2024ztx}. The third column shows the BO quantum numbers  represented as $\left(k\right)_\eta$ instead of $\Lambda^{\sigma}_{\eta}$ , where $\eta = g$ denotes positive parity.}
	\label{tab:QQbarpenta}
\end{table}

\section{Static energies and behavior at short distances}
\label{sec:static}

We consider a bound system of two heavy quarks: $Q\bar{Q}$ or $QQ$ pair along with the LDF (gluons or light quarks). The static energies for quantum numbers $\kappa$ and $\lambda$, $E_{\kappa,|\lambda|}^{\left(0\right)}\left(r\right)$, can be determined  from the large time logarithms of appropriate NRQCD gauge-invariant correlators:
\begin{align}
	E_{\kappa, |\lambda|}^{\left(0\right)}(r)=\lim_{T\to\infty}\frac{i}{T}\,\log\left[ \langle \mathrm{vac}| \mathcal{O}^h_{\kappa, \lambda}(T/2,\,\bm{r},\,\bm{R})\,\mathcal{O}^{h\dagger}_{\kappa, \lambda}(-T/2,\,\bm{r},\,\bm{R})|\mathrm{vac}\rangle\right],\,\, h \in \left\{Q\bar{Q},\,QQ\right\},\label{eq:En}
\end{align}
where $|\mathrm{vac}\rangle$ denotes the NRQCD vacuum and $\mathcal{O}_{\kappa, \lambda}$ is a gauge-invariant interpolating  operator.  For exotic state with either a heavy quark-antiquark octet or a heavy quark-quark antitriplet\footnote{Here, we present results only for the antitriplet and refer to Ref.~\cite{Berwein:2024ztx} for the sextet case}, $\mathcal{O}_{\kappa, \lambda}$ is given by
\begin{align}
	\mathcal{O}^{(Q\bar{Q})_8}_{\kappa, \lambda}\left(t, \bm{r}, \bm{R}\right)&=\nonumber\\
	&\hspace{-1.7 cm}\chi^{\dagger}\left(t, \bm{R}+\bm{r}/2\right)\phi\left(t; \bm{R}+\bm{r}/2,\bm{R}\right)P^{\alpha \dag}_{\kappa, \lambda}H_{8,\,\kappa}^{\alpha,\,a}(t, \bm{R})\,T^a\,\phi\left(t; \bm{R},\bm{R}-\bm{r}/2\right)\psi\left(t, \bm{R}-\bm{r}/2\right),
	\label{eq:intop-1}\\
	\mathcal{O}^{(QQ)_{\bar{3}}}_{\kappa, \lambda}\left(t, \bm{r}, \bm{R}\right)&=\nonumber\\
	&\hspace{-1.7 cm}\psi^T\left(t, \bm{R}+\bm{r}/2\right)\phi^T(t; \bm{R}+\bm{r}/2,\bm{R})P^{\alpha \dag}_{\kappa, \lambda}H_{3,\,\kappa}^{\alpha,\,\ell}\left(t, \bm{R}\right)\underline{T}^\ell\,\phi\left(t; \bm{R}, \bm{R}-\bm{r}/2\right)\psi\left(t,  \bm{R}-\bm{r}/2\right),
	\label{eq:intop-2}
\end{align}  
where  $P^{\alpha}_{\kappa, \lambda}$ are the projectors with $\alpha$ as the vector or spin index, 
 $T^a\,\left(a=1,\cdots,8\right)$ are the $SU(3)$ generators a color octet $\left(Q\bar{Q}\right)_8$ state,  
$\underline{T}^\ell\,\left(\ell = 1, 2, 3\right)$ are generators for color triplet $\left(\bar{Q}\bar{Q}\right)_{3}$ 
or antitriplet $\left(QQ\right)_{\bar{3}}$ state and $\phi\left(\bm{x}, \bm{y}\right)$ denotes the Wilson line.
The operators $H_{8,\,\kappa}^{\alpha,\,a}(\bm{R})$  are related to gluelump operators 
when the LDF are gluons and to adjoint mesons or baryons when the LDF are light quarks~\cite{Foster:1998wu,Michael:1985ne, Campbell:1985kp}.
The operators $H_{3,\,\kappa}^{\alpha,\,\ell}(\bm{R})$  are related to triplet mesons or baryons. Note that, the interpolating operator in Eq.~\eqref{eq:intop-1} with the color octet $\left(Q\bar{Q}\right)_8$ pair has \textit{zero overlap with color-singlet states composed of quarkonium and pions. This avoids a key issue that has complicated some lattice calculations in  the $Q\bar{Q}$ sector} \cite{Prelovsek:2019ywc, Bali:2000vr, Bulava:2019iut, Bulava:2024jpj}. 
Practically, this can aid lattice calculations of the static energies for adjoint $Q\bar{Q}$ tetraquark and pentaquark states, as the interpolating operator in Eq.~\eqref{eq:intop-1} explicitly probes the repulsive color-octet behavior at short distances. We refer to Ref.~\cite{Berwein:2024ztx} for a comprehensive list of the operators  $H_{8,\,\kappa}^{\alpha,\,a}(\bm{R})$ and $H_{3,\,\kappa}^{\alpha,\,\ell}(\bm{R})$ relevant for the static energy (Eq.~\eqref{eq:En}) computations of quarkonium hybrids $\left(Q\bar{Q}g\right)$, quarkonium tetraquarks $\left(Q\bar{Q}q\bar{q}\right)$, quarkonium pentaquarks $\left(Q\bar{Q}qqq\right)$, doubly heavy baryons $\left(QQq\right)$, doubly heavy tetraquarks $\left(QQ\bar{q}\bar{q}\right)$ and doubly heavy pentaquarks $\left(QQqq\bar{q}\right)$. 

Lattice computations of several static energies have been thoroughly explored for hybrids~\cite{Juge:1999ie,Juge:2002br,Bali:2000vr,Bali:2003jq,Capitani:2018rox,Schlosser:2021wnr, Bicudo:2021tsc, Sharifian:2023idc, Hollwieser:2023bud}, while studies for tetraquarks remain relatively sparse~\cite{Brown:2012tm, Bicudo:2015kna, Prelovsek:2019ywc, Sadl:2021bme, Mueller:2023wzd}.

To describe the short-distance behavior $\left(r\to 0\right)$ of the static energies—corresponding to the short-range behavior of  $V^{\left(0\right)}_{\kappa, |\lambda|}$ in Eq.~\eqref{eq:VQQ} ( input in BOEFT coupled Schr\"odinger equations)—we employ weakly coupled pNRQCD. In this limit the inter-quark separation $r$ between heavy quarks satisfies $r \Lambda_{\rm QCD} \ll 1$, implying that the soft scale associated with the relative momentum transfer between the heavy quarks remains perturbative.
In weakly coupled pNRQCD, the degrees of freedom are color singlet $(S)$ and  color octet $(O^a)$ fields for heavy quark-antiquark pairs \cite{Brambilla:1999xf,Brambilla:2004jw}
or color antitriplet $(T^\ell)$ and color sextet $(\Sigma^\sigma)$ fields for heavy quark-quark pairs \cite{Brambilla:2005yk}.

The matching conditions from NRQCD to weaky-coupled pNRQCD to BOEFT for color octet $\left(Q\bar{Q}\right)_8$ and color antriplet $\left(QQ\right)_{\bar{3}}$ cases gives
\begin{align}
	&\mathcal{O}_{\kappa, \lambda}^{(Q\bar{Q})_8}(t,\,\bm{r},\,\bm{R})\longrightarrow \sqrt{Z_{\kappa,\,\lambda}}\, O^a(t,\,\bm{r},\,\bm{R})\,P^{\alpha \dag}_{\kappa, \lambda}\,H^{\alpha, \,a}_{8,\,\kappa}(t,\,\bm{R})+{\cal O}\left(r\right)\longrightarrow \sqrt{Z_{\kappa,\,\lambda}^{Q\bar{Q}}}\,\Psi^{(Q\bar{Q})_8}_{\kappa, \lambda}(t,\,\bm{r},\,\bm{R})\,,
	\label{eq:match-1}\\
	&\mathcal{O}_{\kappa, \lambda}^{(QQ)_{\bar{3}}}(t,\,\bm{r},\,\bm{R}) \longrightarrow \sqrt{Z^{\prime}_{\kappa,\,\lambda}}\,T^\ell(t,\,\bm{r},\,\bm{R})\,P^{\alpha \dag}_{\kappa, \lambda}\,H_{3,\,\kappa}^{\alpha,\,\ell}(t,\,\bm{R})+{\cal O}\left(r\right)\longrightarrow \sqrt{Z_{\kappa,\,\lambda}^{QQ}}\,\Psi^{(QQ)_{\bar{3}}}_{\kappa, \lambda}(t,\,\bm{r},\,\bm{R})\,,
	\label{eq:match-2}
\end{align}
where $Z_{\kappa,\,\lambda}^{h}$, $Z^{\prime}_{\kappa,\,\lambda}$, $Z_{\kappa,\,\lambda}^{Q\bar{Q}}$, and $Z_{\kappa,\,\lambda}^{QQ}$  are field normalization factors (in general a function of ${\bm r}$ and ${\bm R}$) and $\Psi^{(Q\bar{Q})_8}_{\kappa, \lambda}$ and $\Psi^{(QQ)_{\bar{3}}}_{\kappa, \lambda}$ are the BOEFT fields for color octet and color antriplet respectively. 

At short distances, the potentials (static energies) can generally be expressed as the sum of a perturbative component—typically nonanalytic in $r$ and a nonperturbative component coming from the multipole expansion in $r$. The coefficients of the nonperturbative piece scales as $\Lambda_{\rm QCD}$ and can be expressed in terms of gluonic and light-quark correlators. Following Ref.~\cite{Berwein:2024ztx}, the matching condition  for the static
energy computed in NRQCD (left), the short-distance potential in pNRQCD (middle), and
the BOEFT potential (right) gives for the color octet $\left(Q\bar{Q}\right)_8$ case
\begin{align}
	E^{(0)}_{\kappa, |\lambda|}(r) = V_o(r) + \Lambda_{\kappa} + {\cal O}(r^2) = V^{(0)}_{\kappa, |\lambda|}(r).
	\label{eq:QQbarpot-short}
\end{align}
where $V_o(r) = \alpha_s/\left(6r\right)$ is the color octet potential and $\Lambda_{\kappa}$ is either gluelump or adjoint meson or baryon mass correspondig to LDF quantum number $\kappa$ in Eq.~\eqref{label-n}. Note that the static energy and the BOEFT potential are approximated by the middle equality only for small $r$. Eq~\eqref{eq:QQbarpot-short} indicates that, at leading order in the multipole expansion, multiple static energies become degenerate as they depend solely on $\Lambda_{\kappa}$, which has been very clearly seen in the lattice calculations of the hybrid static energies. The degeneracy is broken by the ${\cal O}\left(r^2\right)$ terms. Similarly, in the color anti-triplet $\left(QQ\right)_{\bar{3}}$ case, we get
\begin{align}
	E^{(0)}_{\kappa, |\lambda|}(r) = V_T(r) + \Lambda^T_{\kappa} + {\cal O}(r^2) = V^{(0)}_{\kappa, |\lambda|}(r) \qquad\qquad  ,
	\label{eq:QQpot-short}
\end{align}
where $V_T(r)= -2\alpha_s/\left(3r\right)$ is the color triplet potential and $\Lambda^T_{\kappa}$ is either triplet meson or baryon mass. Again, at leading order in the multipole expansion, multiple static energies will be degenerate. Till date, the lowest adjoint meson spectrum has only been calculated in quenched lattice QCD with light valence quarks \cite{Foster:1998wu} and there are now lattice calculations of the adjoint baryon, triplet meson and triplet baryon masses.

\section{Mixing with heavy-light threshold}
\label{sec:mixing}

The behavior of the static energies at large distances depends on the LDF. In quenched approximation, the LDF are only gluons and the static energies of hybrids and quarkonium increase linearly with $r$ as shown by multiple lattice calculations \cite{Brambilla:2022het,Juge:2002br,Schlosser:2021wnr, Bicudo:2021tsc}. However, in unquenched case, the LDF includes light quarks and  so, we should consider the  static heavy-light threshold states (meson-(anti)meson, baryon-(anti)meson etc). The mixing of the static energies with the heavy-light threshold depends on the BO quantum number $\Lambda_\eta^\sigma$ \cite{Berwein:2024ztx}.

We identify the quantum number $\kappa \equiv \{ k^{P[C]}, f \} $ (Eq.~\eqref{label-n}) for the heavy-light pair states.
Charge conjugation 
$C$ is a good quantum number only for the heavy-light meson-antimeson threshold, while meson-meson or meson-baryon thresholds are characterized by $k$ and parity $P$.
The LDF quantum numbers are inferred from heavy meson and baryon multiplets: the ground state heavy meson $Q\bar{q}$  (antimeson $\bar{Q}q$) has $k^P=(1/2)^-$  ($k^P=(1/2)^+$), followed by two states of similar mass  $k^P=(1/2)^+$ and $k^P=(3/2)^+$ (quark $k^P=(1/2)^-$ and $k^P=(3/2)^-$ ). 
The ground state heavy baryon $Qqq$ corresponds to light quarks $k^P=0^+$ and $k^P=1^+$, 
where we assume that the $k^P=0^+$ state is lower in energy.\footnote{
Our assumption is supported by the observation that $\Lambda$-baryons have lower mass than $\Sigma$-baryons.}
Each static heavy-light threshold state is defined by the spin and parity of the light quark states forming the heavy mesons or baryons. 
Table\ref{tab:qqbar-qq} presents the $k^{PC}$ and BO quanutm number $\Lambda_\eta^\sigma$ for the  $Q\bar{q}$-$\bar{Q}q$ heavy-light threshold states. We refer to ref.~\cite{Berwein:2024ztx} for similar results on $Q\bar{q}$-$Q\bar{q}$, $Qqq$-$\bar{Q}q$ or $Qqq$-$Q\bar{q}$ heavy-light pair states.  Note that in Table~\ref{tab:qqbar-qq}, we considered symmetric and antisymmetric combinations under charge conjugation for the S-wave plus P-wave (shown in second block). The heavy-light pair states also depend on the LDF flavor quantum numbers such as isospin.
\begin{table}[th!]
\begin{center}
\small{\renewcommand{\arraystretch}{0.9}
\scriptsize
\begin{tabular}{|c|c||c|}  \hline\hline
\multirow{2}{*}{\hspace{2pt} $\begin{array}{c} k_{\bar{q}}^{P}\otimes k_q^{P}\end{array} $ \hspace{2pt}} & \multirow{2}{*}{\hspace{2pt} $\begin{array}{c}k^{PC}\end{array} $ \hspace{2pt}} &  \multirow{2}{*}{\hspace{2pt} $\begin{array}{c} \text{BO quantum \#}\\\Lambda^\sigma_\eta\end{array} $ \hspace{2pt}}\\& & \\
\hline\hline
$(1/2)^-\otimes(1/2)^+ $         & $0^{-+}$ & $\Sigma_u^-$ \\
                                & $1^{--}$ & $\Sigma_g^+,\,\Pi_g$ \\ \hline
$(1/2)^-\otimes(1/2)^-\,+\,(1/2)^+\otimes(1/2)^+$         & $0^{+-}$ & $\Sigma_u^+$\\                                 
                                & $1^{+-}$ & $\Sigma_u^-,\,\Pi_u$ \\
$(1/2)^-\otimes(1/2)^-\,-\,(1/2)^+\otimes(1/2)^+$         & $0^{++}$ & $\Sigma_g^+$\\                                 
                                & $1^{++}$ & $\Sigma_g^-,\,\Pi_g$ \\                                
$(1/2)^-\otimes(3/2)^-\,+\,(3/2)^+\otimes(1/2)^+ $      & $1^{+-}$ & $\Sigma_u^-,\,\Pi_u$  \\
                                & $2^{+-}$ & $\Sigma_u^+,\,\Pi_u,\,\Delta_u$\\
$(1/2)^-\otimes(3/2)^-\,-\,(3/2)^+\otimes(1/2)^+$         & $1^{++}$ & $\Sigma_g^-,\,\Pi_g$  \\
                                & $2^{++}$ & $\Sigma_g^+,\,\Pi_g,\,\Delta_g$ \\\hline\hline
\end{tabular}
\caption{The $k^{PC}$ quantum numbers of the light quark-antiquark $\left(\bar{q} q\right)$ pair and  corresponding BO quantum numbers shown in third column for the  $Q\bar{q}$-$\bar{Q}q$ heavy-light threshold states. Each block of states, separated by a single horizontal line, corresponds to S-wave plus S-wave, S-wave plus P-wave. The linear combinations in some entries in the first column refer to symmetric and antisymmetric combination under charge conjugation. 
}
\label{tab:qqbar-qq}}
\end{center}
\end{table}

In $r\rightarrow 0$ limit, weakly-coupled pNRQCD predicts that $Q\bar{Q}$ exotic states exhibit a repulsive octet behavior, as given by Eq.~\eqref{eq:QQbarpot-short}, corresponding to adjoint mesons and baryons. For $QQ$ exotic states, the static energies can be attractive (Eq.~\eqref{eq:QQpot-short}), linked to triplet mesons and baryons, or repulsive, associated with sextet mesons and baryons. \textit{The BO quantum numbers $\Lambda^{\sigma}_{\eta}$ should be conserved at all $r$}. Based on this conservation, the static energies behavior  in the large $r$ region: the quark configurations in $Q\bar{Q}q\bar{q}$ tetraquarks and $Q\bar{Q}qqq$ pentaquarks rearrange to smoothly transition to static heavy-light meson-antimeson $\left(Q\bar{q}\right.$-$\left.\bar{Q}q\right)$ states and static heavy-light baryon-antimeson $\left(Qqq\right.$-$\left.\bar{Q}q\right)$ states respectively. While the quark configurations in $QQ\bar{q}\bar{q}$ tetraquarks and $QQqq\bar{q}$ pentaquarks  rearrange to smoothly transition to static heavy-light meson-meson $\left(Q\bar{q}\right.$-$\left.Q\bar{q}\right)$ states and static heavy-light meson-baryon $\left(Q\bar{q}\right.$-$\left.Qqq\right)$ states.  It was pointed for the first time in \cite{Berwein:2024ztx} and subsequently in \cite{Braaten:2024tbm}. The smooth transition implies no narrow-avoided crossing between the tetraquark and pentaquark static energies and the static heavy-light meson or baryon pair thresholds.

\begin{figure}[ht]
\centering
\begin{minipage}{.45\textwidth}
\centering \includegraphics*[width=\textwidth,clip=true]{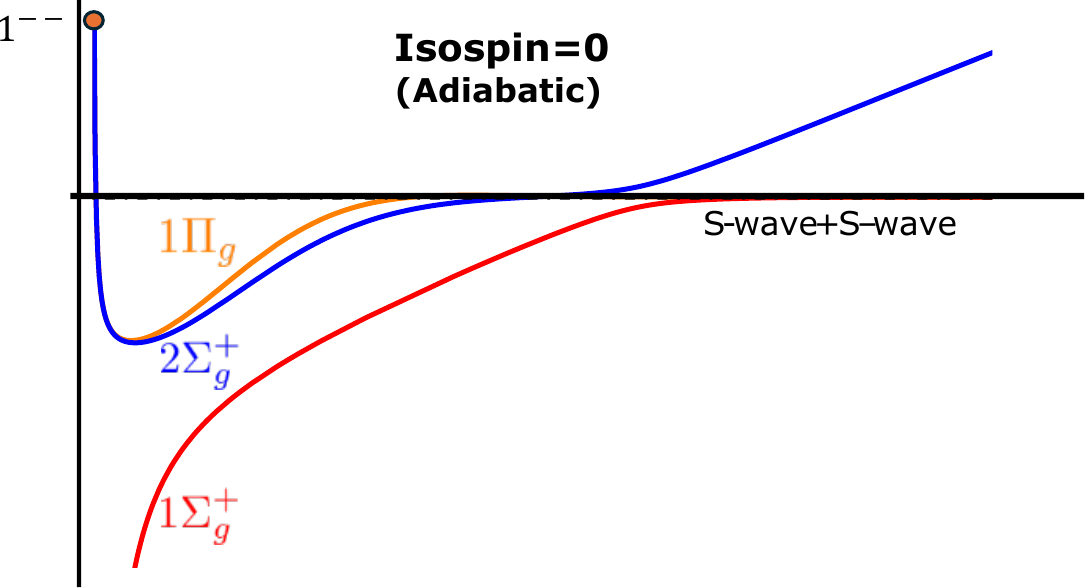}
\end{minipage}
\begin{minipage}{.45\textwidth}
\centering \includegraphics*[width=\textwidth,clip=true]{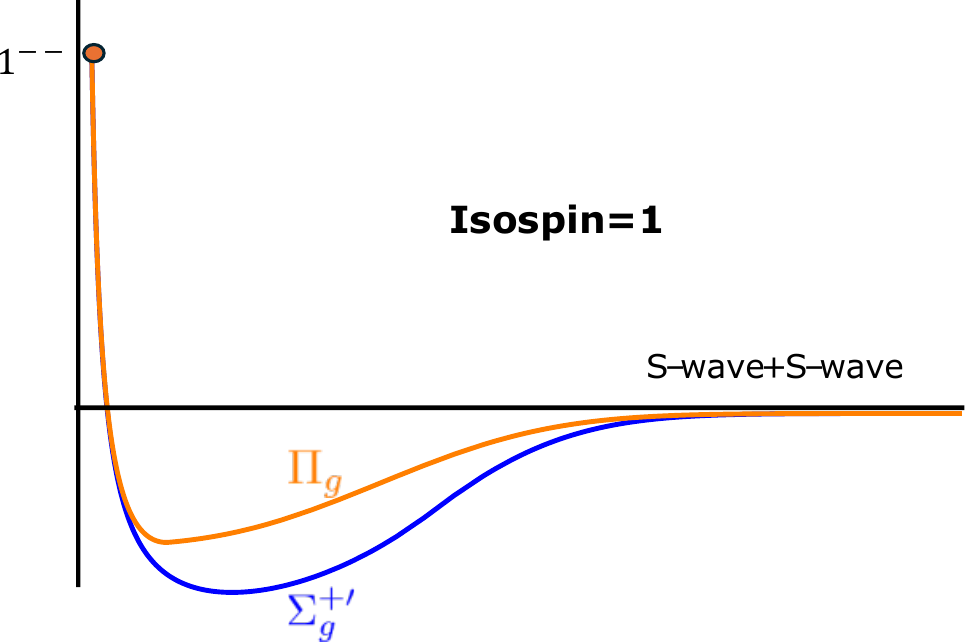}
\end{minipage}
\caption{On the left, we show the lowest $I=0$$Q\bar{Q}$ adiabatic static energies  as functions of $r$~\cite{Berwein:2024ztx}. The avoided levels crossing between the quarkonium potential with   $\Sigma_g^+$ and the  first tetraquark potentials with  $\Sigma_g^{+\prime}$. As $\left(r\rightarrow0\right)$, the adiabatic energy $1\Sigma_g^+$ has attractive behavior corresponding to quarkonium $\Sigma_g^+$ static energy and the adiabatic energies $\{2\Sigma_g^+, 1\Pi_g\}$  have repulsive behavior corresponding to tetraquark static energies $\{\Sigma_g^{+\prime}, \Pi_g\}$  corresponding to $1^{--}$ adjoint meson. The string breaking radius $r_c\sim 1.2$~fm, where there is the avoided level crossing \cite{Bulava:2024jpj, Bali:2000vr, Bulava:2019iut}. On the right, example of a possible  
$I=1$ tetraquark static energy approaching S-wave plus S-wave heavy-light static threshold is shown. Both the $\Pi_g$ and the $\Sigma_g^+$ evolve to the same heavy-light pair that has both BO quantum numbers. 
}
\label{fig:isospin01}
\end{figure}

As an example, this conservation of $\Lambda^{\sigma}_{\eta}$ between small $r$ and large $r$ implies that
in $Q\bar{Q}q\bar{q}$ tetraquarks, the $\Sigma_g^{+\prime}$ and $\Pi_g$ static energies corresponding to adjoint meson $1^{--}$, which have repulsive color octet behavior at small $r$, mix with the S-wave plus S-wave static heavy-light meson-antimeson threshold, which leads to approaching the threshold at large $r$ as shown in Fig.~\ref{fig:isospin01}. A similar mixing occurs for the $\Sigma_u^-$ static energy corresponding to  adjoint meson $0^{-+}$ and the same pattern holds for the $QQ\bar{q}\bar{q}$, $Q\bar{Q}qqq$ and $QQqq\bar{q}$ systems. 

The $I=0$ sector is particularly interesting
as lattice studies have explicitly observed a narrow avoided level crossing between the quarkonium static energy and the heavy-light static energy \cite{Bulava:2024jpj, Bali:2000vr, Bulava:2019iut}.
In our description, this means that two static energies with the same BO quantum numbers, namely
the quarkonium static energy  $\Sigma_g^+$
and the first tetraquark static energy $\Sigma_g^{+\prime}$, become close in region around the string breaking radius $r_c$ with a  non-zero transition amplitude. Near $r_c$, the $\Sigma_g^+$ BO-potential flattens to static heavy-light meson-antimeson pair, while $\Sigma_g^{+\prime}$ continues the long-range confining behavior of quarkonium until the narrow avoided level crossing with another excited tetraquark static-energy with the same $\Lambda^{\sigma}_{\eta}$ (see adiabatic energies in Fig.~\ref{fig:isospin01}) \cite{Berwein:2024ztx}. Note that the $\Pi_g$, which is degenerate with $\Sigma_g^{+\prime}$ as $r\rightarrow 0$ does not show avoided level crossing with $\Sigma_g^+$ near $r_c$ due to their differing $\Lambda^{\sigma}_{\eta}$. Again, similar pattern holds
for $Q\bar{Q}g$ system, where the  narrow avoided level crossing is with S-wave plus P-wave static heavy-light threshold and also in $QQq$ systems, where the narrow avoided level crossing  static heavy-light meson-baryon threshold \cite{Berwein:2024ztx, Bruschini:2024fyj}.

The phenomenological application of the behavior of tetraquark static energies for the experimentally observed states like $\chi_{c1}(3872)$, $Z_b$, and $T_{cc}^+\left(3875\right)$ has been explored in Refs.~\cite{Berwein:2024ztx, Braaten:2024tbm}.

\section{Conclusions}

In this work, we have developed a QCD-based description of states containing two heavy quarks—$Q\bar{Q}$ or $QQ$—including exotic hybrids, tetraquarks, and pentaquarks. Our approach is grounded in the Born–Oppenheimer nonrelativistic effective field theory (BOEFT), which systematically factorizes the heavy quark dynamics from the light degrees of freedom (LDF) based on the hierarchy $\Lambda_{\mathrm{QCD}}\gg E$. In Sec.~\ref{sec:Sch}, we derived a general expression for the coupled Schr\"odinger equations for an arbitrary LDF quantum number. These coupled equations are new in the context of tetraquarks and pentaquarks, with explicit forms for integer-spin (hybrids, tetraquarks) and half-integer-spin (doubly heavy baryons, pentaquarks) states given in Ref.~\cite{Berwein:2024ztx}. In Sec.~\ref{sec:static}, we express the static energies for $Q\bar{Q}$ and $QQ$ exotic states in terms of gauge-invariant correlators computable on the lattice and analyze their short-distance behavior through matching to pNRQCD. In $r\rightarrow 0$ limit, he static energy depends on just two nonperturbative parameters: the mass-dimension-one constant $\Lambda_\kappa$ and the quadratic slope, as in Eqs.~\eqref{eq:QQbarpot-short} and \eqref{eq:QQpot-short}. In Sec.~\ref{sec:mixing}, we examine the long-distance behavior of exotic static energies, particularly their mixing with the static heavy-light pair threshold. This mixing occurs only if both share the same BO quantum numbers $\Lambda^\sigma_\eta$.  A key conclusion is that for tetraquark and pentaquark static energies, this mixing proceeds without string breaking—a novel finding.

\begin{acknowledgments}
We thank Nora Brambilla, Mathias Berwein, and Antonio Vairo for advice and collaboration on this work. This work has been supported by the DFG cluster of excellence ORIGINS funded by the Deutsche Forschungsgemeinschaft under Germany’s Excellence Strategy-EXC-2094-390783311 and the STRONG-2020-Strong Interaction at the Frontier of Knowledge: Fundamental Research and Applications”. STRONG-2020 has received funding from the European
Union’s Horizon 2020 research and innovation program under grant agreement No. 824093.
We also acknowledge the European Union ERC-2023-ADG-Project EFT-XYZ.  
\end{acknowledgments}

\bibliographystyle{utphys-mod3}
\bibliography{One-BOEFT.bib}

\providecommand{\href}[2]{#2}\begingroup\raggedright\begin{thebibliography}{10}
  \setlength{\itemsep}{0mm}

\bibitem{GellMann:1964nj}
M.~Gell-Mann,
\href{http://dx.doi.org/10.1016/S0031-9163(64)92001-3}{{\em Phys. Lett.} {\bf
  8} (1964)  214--215}

\bibitem{Zweig:1964CERN}
G.~Zweig, {\em CERN Report No.8182/TH.401, CERN Report No.8419/TH.412} (1964)

\bibitem{Jaffe:1975fd}
R.~L. Jaffe and K.~Johnson,
  \href{http://dx.doi.org/10.1016/0370-2693(76)90423-8}{{\em Phys. Lett. B}
  {\bf 60} (1976)  201--204}

\bibitem{Belle:2003nnu}
{\bf Belle} Collaboration: S.~K. Choi {\em et al.},
  \href{http://dx.doi.org/10.1103/PhysRevLett.91.262001}{{\em Phys. Rev. Lett.}
  {\bf 91} (2003)  262001}

\bibitem{Guo:2017jvc}
F.-K. Guo, C.~Hanhart, U.-G. Mei\ss{}ner, Q.~Wang, Q.~Zhao, and B.-S. Zou,
  \href{http://dx.doi.org/10.1103/RevModPhys.90.015004}{{\em Rev. Mod. Phys.}
  {\bf 90} (2018) no.~1, 015004}

\bibitem{Ali:2017jda}
A.~Ali, J.~S. Lange, and S.~Stone,
  \href{http://dx.doi.org/10.1016/j.ppnp.2017.08.003}{{\em Prog. Part. Nucl.
  Phys.} {\bf 97} (2017)  123--198}

\bibitem{Olsen:2017bmm}
S.~L. Olsen, T.~Skwarnicki, and D.~Zieminska,
  \href{http://dx.doi.org/10.1103/RevModPhys.90.015003}{{\em Rev. Mod. Phys.}
  {\bf 90} (2018) no.~1, 015003}

\bibitem{Brambilla:2019esw}
N.~Brambilla, S.~Eidelman, C.~Hanhart, A.~Nefediev, C.-P. Shen, C.~E. Thomas,
  A.~Vairo, and C.-Z. Yuan,
  \href{http://dx.doi.org/10.1016/j.physrep.2020.05.001}{{\em Phys. Rept.} {\bf
  873} (2020)  1--154}

\bibitem{Liu:2019zoy}
Y.-R. Liu, H.-X. Chen, W.~Chen, X.~Liu, and S.-L. Zhu,
  \href{http://dx.doi.org/10.1016/j.ppnp.2019.04.003}{{\em Prog. Part. Nucl.
  Phys.} {\bf 107} (2019)  237--320}

\bibitem{Chen:2022asf}
H.-X. Chen, W.~Chen, X.~Liu, Y.-R. Liu, and S.-L. Zhu,
  \href{http://dx.doi.org/10.1088/1361-6633/aca3b6}{{\em Rept. Prog. Phys.}
  {\bf 86} (2023) no.~2, 026201}

\bibitem{LHCb:2015yax}
{\bf LHCb} Collaboration: R.~Aaij {\em et al.},
  \href{http://dx.doi.org/10.1103/PhysRevLett.115.072001}{{\em Phys. Rev.
  Lett.} {\bf 115} (2015)  072001}

\bibitem{LHCb:2019kea}
{\bf LHCb} Collaboration: R.~Aaij {\em et al.},
  \href{http://dx.doi.org/10.1103/PhysRevLett.122.222001}{{\em Phys. Rev.
  Lett.} {\bf 122} (2019) no.~22, 222001}

\bibitem{Berwein:2024ztx}
M.~Berwein, N.~Brambilla, A.~Mohapatra, and A.~Vairo,
  \href{http://dx.doi.org/10.1103/PhysRevD.110.094040}{{\em Phys. Rev. D} {\bf
  110} (2024) no.~9, 094040}

\bibitem{Caswell:1985ui}
W.~E. Caswell and G.~P. Lepage,
  \href{http://dx.doi.org/10.1016/0370-2693(86)91297-9}{{\em Phys. Lett. B}
  {\bf 167} (1986)  437--442}

\bibitem{Bodwin:1994jh}
G.~T. Bodwin, E.~Braaten, and G.~P. Lepage,
  \href{http://dx.doi.org/10.1103/PhysRevD.55.5853}{{\em Phys. Rev. D} {\bf 51}
  (1995)  1125--1171}

\bibitem{Manohar:1997qy}
A.~V. Manohar, \href{http://dx.doi.org/10.1103/PhysRevD.56.230}{{\em Phys. Rev.
  D} {\bf 56} (1997)  230--237}

\bibitem{Brambilla:2004jw}
N.~Brambilla, A.~Pineda, J.~Soto, and A.~Vairo,
  \href{http://dx.doi.org/10.1103/RevModPhys.77.1423}{{\em Rev. Mod. Phys.}
  {\bf 77} (2005)  1423}

\bibitem{Berwein:2015vca}
M.~Berwein, N.~Brambilla, J.~Tarr\'us~Castell\`a, and A.~Vairo,
  \href{http://dx.doi.org/10.1103/PhysRevD.92.114019}{{\em Phys. Rev. D} {\bf
  92} (2015) no.~11, 114019}

\bibitem{Oncala:2017hop}
R.~Oncala and J.~Soto, \href{http://dx.doi.org/10.1103/PhysRevD.96.014004}{{\em
  Phys. Rev. D} {\bf 96} (2017) no.~1, 014004}

\bibitem{Brambilla:2017uyf}
N.~Brambilla, G.~a. Krein, J.~Tarr\'us~Castell\`a, and A.~Vairo,
  \href{http://dx.doi.org/10.1103/PhysRevD.97.016016}{{\em Phys. Rev. D} {\bf
  97} (2018) no.~1, 016016}

\bibitem{Brambilla:1999xf}
N.~Brambilla, A.~Pineda, J.~Soto, and A.~Vairo,
  \href{http://dx.doi.org/10.1016/S0550-3213(99)00693-8}{{\em Nucl. Phys. B}
  {\bf 566} (2000)  275}

\bibitem{Pineda:1997bj}
A.~Pineda and J.~Soto,
  \href{http://dx.doi.org/10.1016/S0920-5632(97)01102-X}{{\em Nucl. Phys. B
  Proc. Suppl.} {\bf 64} (1998)  428--432}

\bibitem{Brambilla:2000gk}
N.~Brambilla, A.~Pineda, J.~Soto, and A.~Vairo,
  \href{http://dx.doi.org/10.1103/PhysRevD.63.014023}{{\em Phys. Rev. D} {\bf
  63} (2001)  014023}

\bibitem{Pineda:2000sz}
A.~Pineda and A.~Vairo,
  \href{http://dx.doi.org/10.1103/PhysRevD.64.039902}{{\em Phys. Rev. D} {\bf
  63} (2001)  054007}

\bibitem{Born-Oppenheimer}
M.~Born and R.~Oppenheimer,
  \href{http://dx.doi.org/https://doi.org/10.1002/andp.19273892002}{{\em
  Annalen der Physik} {\bf 389} (1927) no.~20, 457--484}

\bibitem{Landau:1991wop}
L.~D. Landau and E.~M. Lifshits, {\em {Quantum Mechanics}: {Non-Relativistic
  Theory}} vol.~v.3 of {\em Course of Theoretical Physics}.
\newblock Butterworth-Heinemann Oxford 1991

\bibitem{Griffiths:1983ah}
L.~A. Griffiths, C.~Michael, and P.~E.~L. Rakow,
  \href{http://dx.doi.org/10.1016/0370-2693(83)90680-9}{{\em Phys. Lett. B}
  {\bf 129} (1983)  351--356}

\bibitem{Juge:1999ie}
K.~J. Juge, J.~Kuti, and C.~J. Morningstar,
  \href{http://dx.doi.org/10.1103/PhysRevLett.82.4400}{{\em Phys. Rev. Lett.}
  {\bf 82} (1999)  4400--4403}

\bibitem{Juge:2002br}
K.~J. Juge, J.~Kuti, and C.~Morningstar,
  \href{http://dx.doi.org/10.1103/PhysRevLett.90.161601}{{\em Phys. Rev. Lett.}
  {\bf 90} (2003)  161601}

\bibitem{Braaten:2013boa}
E.~Braaten, \href{http://dx.doi.org/10.1103/PhysRevLett.111.162003}{{\em Phys.
  Rev. Lett.} {\bf 111} (2013)  162003}

\bibitem{Braaten:2014qka}
E.~Braaten, C.~Langmack, and D.~H. Smith,
  \href{http://dx.doi.org/10.1103/PhysRevD.90.014044}{{\em Phys. Rev. D} {\bf
  90} (2014) no.~1, 014044}

\bibitem{Braaten:2014ita}
E.~Braaten, C.~Langmack, and D.~H. Smith,
  \href{http://dx.doi.org/10.1103/PhysRevLett.112.222001}{{\em Phys. Rev.
  Lett.} {\bf 112} (2014)  222001}

\bibitem{Meyer:2015eta}
C.~A. Meyer and E.~S. Swanson,
  \href{http://dx.doi.org/10.1016/j.ppnp.2015.03.001}{{\em Prog. Part. Nucl.
  Phys.} {\bf 82} (2015)  21--58}

\bibitem{Alasiri:2024nue}
F.~Alasiri, E.~Braaten, and A.~Mohapatra,
  \href{http://dx.doi.org/10.1103/PhysRevD.110.054029}{{\em Phys. Rev. D} {\bf
  110} (2024) no.~5, 054029}

\bibitem{Brambilla:2022hhi}
N.~Brambilla, W.~K. Lai, A.~Mohapatra, and A.~Vairo,
  \href{http://dx.doi.org/10.1103/PhysRevD.107.054034}{{\em Phys. Rev. D} {\bf
  107} (2023) no.~5, 054034}

\bibitem{TarrusCastella:2021pld}
J.~Tarr\'us~Castell\`a and E.~Passemar,
  \href{http://dx.doi.org/10.1103/PhysRevD.104.034019}{{\em Phys. Rev. D} {\bf
  104} (2021) no.~3, 034019}

\bibitem{Brambilla:2018pyn}
N.~Brambilla, W.~K. Lai, J.~Segovia, J.~Tarr\'us~Castell\`a, and A.~Vairo,
  \href{http://dx.doi.org/10.1103/PhysRevD.99.014017}{{\em Phys. Rev. D} {\bf
  99} (2019) no.~1, 014017}

\bibitem{Brambilla:2019jfi}
N.~Brambilla, W.~K. Lai, J.~Segovia, and J.~Tarr\'us~Castell\`a,
  \href{http://dx.doi.org/10.1103/PhysRevD.101.054040}{{\em Phys. Rev. D} {\bf
  101} (2020) no.~5, 054040}

\bibitem{Soto:2023lbh}
J.~Soto and S.~T. Valls,
  \href{http://dx.doi.org/10.1103/PhysRevD.108.014025}{{\em Phys. Rev. D} {\bf
  108} (2023) no.~1, 014025}

\bibitem{Soto:2020xpm}
J.~Soto and J.~Tarr\'us~Castell\`a,
  \href{http://dx.doi.org/10.1103/PhysRevD.102.014012}{{\em Phys. Rev. D} {\bf
  102} (2020) no.~1, 014012}

\bibitem{Soto:2020pfa}
J.~Soto and J.~Tarr\'us~Castell\`a,
  \href{http://dx.doi.org/10.1103/PhysRevD.104.059901}{{\em Phys. Rev. D} {\bf
  102} (2020) no.~1, 014013}

\bibitem{Soto:2021cgk}
J.~Soto and J.~Tarr\'us~Castell\`a,
  \href{http://dx.doi.org/10.1103/PhysRevD.104.074027}{{\em Phys. Rev. D} {\bf
  104} (2021)  074027}

\bibitem{Brambilla:2005yk}
N.~Brambilla, A.~Vairo, and T.~R{\"o}sch,
  \href{http://dx.doi.org/10.1103/PhysRevD.72.034021}{{\em Phys. Rev. D} {\bf
  72} (2005)  034021}

\bibitem{Bicudo:2015kna}
P.~Bicudo, K.~Cichy, A.~Peters, and M.~Wagner,
  \href{http://dx.doi.org/10.1103/PhysRevD.93.034501}{{\em Phys. Rev. D} {\bf
  93} (2016) no.~3, 034501}

\bibitem{Bicudo:2016ooe}
P.~Bicudo, J.~Scheunert, and M.~Wagner,
  \href{http://dx.doi.org/10.1103/PhysRevD.95.034502}{{\em Phys. Rev. D} {\bf
  95} (2017) no.~3, 034502}

\bibitem{Bicudo:2017szl}
P.~Bicudo, M.~Cardoso, A.~Peters, M.~Pflaumer, and M.~Wagner,
  \href{http://dx.doi.org/10.1103/PhysRevD.96.054510}{{\em Phys. Rev. D} {\bf
  96} (2017) no.~5, 054510}

\bibitem{Bicudo:2019ymo}
P.~Bicudo, M.~Cardoso, N.~Cardoso, and M.~Wagner,
  \href{http://dx.doi.org/10.1103/PhysRevD.101.034503}{{\em Phys. Rev. D} {\bf
  101} (2020) no.~3, 034503}

\bibitem{Braaten:2020nwp}
E.~Braaten, L.-P. He, and A.~Mohapatra,
  \href{http://dx.doi.org/10.1103/PhysRevD.103.016001}{{\em Phys. Rev. D} {\bf
  103} (2021) no.~1, 016001}

\bibitem{Brambilla:2024thx}
N.~Brambilla, A.~Mohapatra, T.~Scirpa, and A.~Vairo,
  \href{http://arxiv.org/abs/2411.14306}{{\tt 2411.14306[hep-ph]}}

\bibitem{Braaten:2024tbm}
E.~Braaten and R.~Bruschini,
  \href{http://dx.doi.org/10.1016/j.physletb.2025.139386}{{\em Phys. Lett. B}
  {\bf 863} (2025)  139386}

\bibitem{Bruschini:2024fyj}
R.~Bruschini, \href{http://dx.doi.org/10.1103/PhysRevD.110.074033}{{\em Phys.
  Rev. D} {\bf 110} (2024) no.~7, 074033}

\bibitem{Hoffmann:2024hbz}
J.~Hoffmann and M.~Wagner,
  \href{http://dx.doi.org/10.1103/PhysRevD.111.054507}{{\em Phys. Rev. D} {\bf
  111} (2025) no.~5, 054507}

\bibitem{TarrusCastella:2022rxb}
J.~Tarr\'us~Castell\`a,
  \href{http://dx.doi.org/10.1103/PhysRevD.106.094020}{{\em Phys. Rev. D} {\bf
  106} (2022) no.~9, 094020}

\bibitem{Bruschini:2023zkb}
R.~Bruschini, \href{http://dx.doi.org/10.1007/JHEP08(2023)219}{{\em JHEP} {\bf
  08} (2023)  219}

\bibitem{Bruschini:2023tmm}
R.~Bruschini, \href{http://dx.doi.org/10.1103/PhysRevD.109.L031501}{{\em Phys.
  Rev. D} {\bf 109} (2024) no.~3, L031501}

\bibitem{TarrusCastella:2024zps}
J.~Tarr\'us~Castell\`a, \href{http://dx.doi.org/10.1007/JHEP06(2024)107}{{\em
  JHEP} {\bf 06} (2024)  107}

\bibitem{Braaten:2024stn}
E.~Braaten and R.~Bruschini,
  \href{http://dx.doi.org/10.1103/PhysRevD.109.094051}{{\em Phys. Rev. D} {\bf
  109} (2024) no.~9, 094051}

\bibitem{Alberti:2016dru}
M.~Alberti, G.~S. Bali, S.~Collins, F.~Knechtli, G.~Moir, and W.~S\"oldner,
  \href{http://dx.doi.org/10.1103/PhysRevD.95.074501}{{\em Phys. Rev. D} {\bf
  95} (2017) no.~7, 074501}

\bibitem{Prelovsek:2019ywc}
S.~Prelovsek, H.~Bahtiyar, and J.~Petkovic,
  \href{http://dx.doi.org/10.1016/j.physletb.2020.135467}{{\em Phys. Lett. B}
  {\bf 805} (2020)  135467}

\bibitem{Foster:1998wu}
{\bf UKQCD} Collaboration: M.~Foster and C.~Michael,
  \href{http://dx.doi.org/10.1103/PhysRevD.59.094509}{{\em Phys. Rev. D} {\bf
  59} (1999)  094509}

\bibitem{Michael:1985ne}
C.~Michael, \href{http://dx.doi.org/10.1016/0550-3213(85)90297-4}{{\em Nucl.
  Phys. B} {\bf 259} (1985)  58--76}

\bibitem{Campbell:1985kp}
N.~A. Campbell, I.~H. Jorysz, and C.~Michael,
  \href{http://dx.doi.org/10.1016/0370-2693(86)90552-6}{{\em Phys. Lett. B}
  {\bf 167} (1986)  91--93}

\bibitem{Bali:2000vr}
{\bf TXL, T(X)L} Collaboration: G.~S. Bali, B.~Bolder, N.~Eicker, T.~Lippert,
  B.~Orth, P.~Ueberholz, K.~Schilling, and T.~Struckmann,
  \href{http://dx.doi.org/10.1103/PhysRevD.62.054503}{{\em Phys. Rev. D} {\bf
  62} (2000)  054503}

\bibitem{Bulava:2019iut}
J.~Bulava, B.~H\"orz, F.~Knechtli, V.~Koch, G.~Moir, C.~Morningstar, and
  M.~Peardon, \href{http://dx.doi.org/10.1016/j.physletb.2019.05.018}{{\em
  Phys. Lett. B} {\bf 793} (2019)  493--498}

\bibitem{Bulava:2024jpj}
J.~Bulava, F.~Knechtli, V.~Koch, C.~Morningstar, and M.~Peardon,
  \href{http://dx.doi.org/10.1016/j.physletb.2024.138754}{{\em Phys. Lett. B}
  {\bf 854} (2024)  138754}

\bibitem{Bali:2003jq}
G.~S. Bali and A.~Pineda,
  \href{http://dx.doi.org/10.1103/PhysRevD.69.094001}{{\em Phys. Rev. D} {\bf
  69} (2004)  094001}

\bibitem{Capitani:2018rox}
S.~Capitani, O.~Philipsen, C.~Reisinger, C.~Riehl, and M.~Wagner,
  \href{http://dx.doi.org/10.1103/PhysRevD.99.034502}{{\em Phys. Rev. D} {\bf
  99} (2019) no.~3, 034502}

\bibitem{Schlosser:2021wnr}
C.~Schlosser and M.~Wagner,
  \href{http://dx.doi.org/10.1103/PhysRevD.105.054503}{{\em Phys. Rev. D} {\bf
  105} (2022) no.~5, 054503}

\bibitem{Bicudo:2021tsc}
P.~Bicudo, N.~Cardoso, and A.~Sharifian,
  \href{http://dx.doi.org/10.1103/PhysRevD.104.054512}{{\em Phys. Rev. D} {\bf
  104} (2021) no.~5, 054512}

\bibitem{Sharifian:2023idc}
A.~Sharifian, N.~Cardoso, and P.~Bicudo,
  \href{http://dx.doi.org/10.1103/PhysRevD.107.114507}{{\em Phys. Rev. D} {\bf
  107} (2023) no.~11, 114507}

\bibitem{Hollwieser:2023bud}
R.~H\"ollwieser, F.~Knechtli, T.~Korzec, M.~Peardon, and J.~A. Urrea-Ni\~no,
  ``{Hybrid static potentials from Laplacian Eigenmodes},''
\newblock 12, 2023.
\newblock \href{http://arxiv.org/abs/2401.09453}{{\tt 2401.09453[hep-lat]}}

\bibitem{Brown:2012tm}
Z.~S. Brown and K.~Orginos,
  \href{http://dx.doi.org/10.1103/PhysRevD.86.114506}{{\em Phys. Rev. D} {\bf
  86} (2012)  114506}

\bibitem{Sadl:2021bme}
M.~Sadl and S.~Prelovsek,
  \href{http://dx.doi.org/10.1103/PhysRevD.104.114503}{{\em Phys. Rev. D} {\bf
  104} (2021) no.~11, 114503}

\bibitem{Mueller:2023wzd}
L.~Mueller, P.~Bicudo, M.~Krstic~Marinkovic, and M.~Wagner,
  ``{Antistatic-antistatic-light-light potentials from lattice QCD},''
\newblock 12, 2023.
\newblock \href{http://arxiv.org/abs/2312.17060}{{\tt 2312.17060[hep-lat]}}

\bibitem{Brambilla:2022het}
{\bf TUMQCD} Collaboration: N.~Brambilla, R.~L. Delgado, A.~S. Kronfeld,
  V.~Leino, P.~Petreczky, S.~Steinbei\ss{}er, A.~Vairo, and J.~H. Weber,
  \href{http://dx.doi.org/10.1103/PhysRevD.107.074503}{{\em Phys. Rev. D} {\bf
  107} (2023) no.~7, 074503}

\end{thebibliography}\endgroup



\end{document}